\documentclass[a4paper,10pt]{article}
\usepackage[utf8]{inputenc}
\usepackage[style=ieee,backend=biber]{biblatex}
\usepackage{graphicx}
\usepackage{longtable}
\usepackage{capt-of}
\bibliography{biblo}
\title{Interactive Complexity: Software Metrics from an Ecosystem Perspective}
\author{Charles Hathaway \\ Ron Eglash \\ Mukkai Krishnamoorthy \\ Department of Computer Science\\ Rensselaer Polytechnic Institute, Troy, NY}

\begin{document}

\maketitle
\begin{abstract}
With even the most trivial of applications now being written on top of millions of lines code of libraries, API's, and programming languages, much of the complexity that used to exist when designing software has been abstracted away to allow programmers to focus on primarily business logic.
With each application relying so heavily on the ecosystem it was designed to run in, whether that is limited to a local system or includes dependencies on machines connected by networks, measuring the complexity of these systems can no longer be done simply by observing the code internal to the application; we also need to account for its external interactions.
This is especially important when considering issues of security, which becomes more vital as our healthcare, financial, and automobiles rely on complicated software systems.
We propose Interactive Complexity, which provide a quantitative measure of how intertwined parts of the system are.
Some of the most well-known software complexity metrics out there are the metrics in the CK-metric suite\cite{chidamber1994metrics}; these metrics are designed for use in measuring object oriented systems, but we believe they can be adapted to help measure the interaction of software systems.
Our experimental results show strong correlations between the number of bugs fixed in a release and the value of some of these metrics in systems of sufficient scale.

\end{abstract}

\section{Introduction}

Despite many years of development and contemplation, only a handful of software metrics have succeeded in becoming more than just a glint in the eye of their creators \cite{fenton_software_1999}.
Of the few metrics that made it to fruition, the most successful has been, without a doubt, the CK-metric suite proposed to measure the quality of object oriented systems \cite{tang_empirical_1999} \cite{subramanyam_empirical_2003}.
Rather than just being formulated with the general intent of measuring the complexity of a piece of software, they were designed to help predict the cost of maintaining the software over a period of time; in essence, this is a more practical definition of software quality.
In this paper, we aim to take these tried and tested metrics and extend them to a broader domain: instead of looking at a piece of software as a standalone entity, whether it be composed of classes and objects or anonymous functions and monads, we would like to look at the ecosystem upon which it is built.

Previous works that examined software complexity have focused on a variety of goals.
 McCabe was looking for a metric to quantify the number of testing paths within the code, his belief being that reducing the number of code paths that needed to be tested would increase test coverage.
Others \cite{munson1991use} have been concerned with software reliability.
Another common usage has been in quantifying the cost of maintaining the software over a period of time; typically by an indicator of the probability of faults that a software project would have.
In this paper, we explore the latter, as it has been more widely adopted by the previous literature and is easier to measure.

It is increasingly common in literature on networks to compare software systems to biological systems \cite{mens2014studying}.
This analogy is very helpful in thinking about a metric  for interactive complexity.
Just as ecologists draw a portrait of an organism’s properties by looking at its relations to the surrounding ecosystem , we need to look at a software project’s relation to the digital networks in which it is embedded into get a better understanding of its complexity.
Within a biological system, one needs to examine how an organism interacts with flows of biotic substances and resources:  does it provide a food source? Make homes for others, like beavers or coral? Convert nitrogen in the air to nitrogen in the soil? Pass along viruses? Does it consume a great deal of other resources within the system?

One way ecologists summarize these distinctions is the contrast between “keystone species”--organisms that have so many crucial outgoing links that the ecosystem would be drastically affected if they vanished--and “umbrella species” that take in a great variety of resources but do not have much effect if they vanish \cite{mills1993keystone} \cite{roberge2004usefulness}.
The honeybee is a good example of a keystone species: it is internally simple, but it pollinates hundreds of different plants, upon whom thousands of other species depend.
At the other end of the spectrum, we have umbrella species such as the Siberian tiger: internally far more complex than a bee, but its ecosystem relationships are primarily incoming links, spread over a very large range (it it would make little change to the ecosystem if it vanishes).

We believe that software developed in this hyper-connected era also exists on such a spectrum of interactivity between keystone and umbrella species.
Some software systems are simple, but have huge impacts on the environment (for example, JUnit), and other systems depend on huge numbers of libraries to function (for example, Apache Ambari).
These are the bees and tigers, respectively.

Understanding where in the spectrum between keystone species and umbrella species a particular software project might lie provides some useful insights.
For example, Java Applets were a huge part of the early Internet's success; whether it be small games, VPN's, or productivity applications, Java applets were a solid basis for making the application due to its ease of deployment and versatility.
And yet, if we go look at the Internet today, these applets have all been replaced; why?
To get the answer, all you need to do is find a Java applet, and run it.
You will first have to get past several warning messages about how dangerous this is for your computer, about how there are known vulnerabilities, and how your world will end should you run this Java applet.
This is a case of an keystone species becoming endangered, and resulting in the demise of all umbrella species which relied on it.

Developing software products is complex, and full of intricacies.
However, we summarize it quite simply as the following two activities:
\begin{enumerate}
  \item Write software to solve a problem
  \item Find and use a library or tool that solves a problem
\end{enumerate}

If you write the code from scratch, you can rest assured that it will continue to work as you wrote for a good long while (at least until the underlying system architecture changes).
However, if you write your code on top of a library, the moment the library gets and update, your code may suddenly cease to function.
We hope that Interactive Complexity might help us to find the sweet spot; the place where
using libraries reduces the initial complexity of software, but doesn't caused us to enter “dependency hell”.

Below, we discuss what interactive complexity is and where it comes from, including how we derived these definitions from the CK-metric suite.
We then discuss our case study, done with the Software Metrics Framework \cite{software_metrics_framework}, which provides tools help rerun this study on new metrics and ideas.
Lastly, we examine the results of our case study, including examining some key projects which we believe represent a good sample of the things we indexed.

It is worth noting that examining how software communities are connected has been of great interest in recent years.
Batista et al have recently published work examining the way developer communities on Github interact, and how these interactions build teams \cite{batista_collaboration_2017}.
In many ways, the work we present here is related, with the key distinction being that we are looking at the software community instead of the developer community.

\section{Adapting CK-Metrics to Software Systems}

The CK-metric suite consists of 6 metrics geared towards object-oriented programming systems.

We would like to borrow from the work done by Chidamber and Kemerer \cite{chidamber1994metrics}, and apply their logic to the problem of software systems rather than an individual piece of software.
In this section, we will examine the original intent and definition of each CK-metric, then discuss our new definition and the reasoning behind each non-obvious decision.

Some of their metrics already explored the use of libraries, if not intentionally.
Namely, the DIT (depth of inheritance tree) metric, which would examine the list of classes that an application class was built on top of.
Aside from this metric, none of the others explicitly looked at classes declared in a library; if they happened upon one while calculating the value for an application class, it may perform some metric calculations to help resolve the application class metrics, but it wouldn't continue to explore the library.
For the IC-metric suite, all of these metrics will be re-purposed; even if they had this tangential examination of related libraries.

\subsection{Defining libraries and dependencies}

Rather than briefly mention this question in the introduction, we will delve into what we consider a library, since there are some nuances that need to be considered.
The most obvious definition would be something with the word "library" in the description, such as a library to parse HTML or a library to interface with an Application Program Interface (API).
These types of libraries are usually designed as part of the development of a project to fulfill a need in some way.
As such, they don't directly interface with the end-user (this is usually the domain of another module in the project).

One of the first problems to arise with that definition is that it doesn't account for facilities built into the programming language or operating system itself.
For example, is a call to the C function \texttt{sizeof} considered a library call?
What about a call to \texttt{Enum.reduce} function in Elixir?
\texttt{System.out.println} in Java?

There is also the case of a call to an external system; suppose one wants to get a list of stock prices, and the stock broker has a system failure.
That failure will be reflected in the calling application, even though we didn't directly include the API in our list of libraries.
We would have a record of including a REST (REpresentational State Transfer) client, but we have no way of knowing if that client is used just to access the stock broker API, or that it is used to access more than one API.

These types of questions are what led us away from simply making a metric which looked at the metadata surrounding a project (such as its \texttt{<dependencies>} tag in Maven) or examine bytecode to find symbols (as recent work as has done \cite{lutellier2017measuring}), and towards a metric suite which examined more than one definition of library.
For us, a library is any external dependency; whether it be operating system code, programming language features, third-party libraries, external systems, or modules loaded at run time.
Our goal is to have something that can account for each of these dependencies in some way, and examine how these dependencies fit into the rest of the system.
To that end, we defined the following 6 metrics.

\subsection{Weighted Methods per Class (WMC)}

Chidamber and Kemerer originally thought of this metric as the number of methods per class times their complexity; however, since there was (and still is not) a good metric for determining complexity of a method, this metric is usually considered the count of methods per class.

In our iteration of these metrics, it helps to conceptualize a program as a class; with that scaffolding, we at least have a good avenue to ask questions.
One could imagine it being the number of classes in the program, but since we aren't truly interested in how this program interacts with itself, this isn't of much interest to us.
Instead, if we imagine a software system as a collection of components, where each component is an included library, we could envision mapping this to something like the sum of methods in the included libraries.

Our concern with doing the sum of methods in included libraries is related to the six degrees of seperation concept; most projects will include one or two key libraries.
These libraries will likely include a lot of functionality, and so will artificially bloat the WMC score of projects.
With this in mind, we opted to take the number of dependant libraries as a metric, which is both an obvious and simple metric.
This is measured primarily using the metadata available in Maven, since we do not wish to count libraries multiple times if they are included in many files.

\subsection{Depth of Inheritance Tree (DIT)}

As discussed, this was meant to be the number of parents until a class reaches the base object; the last item with no parents.
We can easily map this to the depth of the dependency tree; keep looking at parents until we find a library that doesn't have a parent, and take the length of the longest chain.

Originally, we had thought of a dependency as any project and versions that was required by something; however, after some investigation we found that including the version as a criteria for selecting the parent resulted in exponential growth of the number of parents we needed to explore.
We briefly considered using concepts from 6-degrees of separation to help limit this growth, but after we stopped using versions as a qualifier, the growth was much more acceptable.

In the end, we ended up using the dependency plugin in Maven to construct a dependency tree, then took the maximum depth of that tree to be the DIT measure.
This was much simplier than initial efforts, which included dumping the list of dependencies across project (which often included multiple libraries) and then recursing on each one, until we found a project with no dependencies.

\subsection{Number of Children (NOC)}


Originally defined as the number of children that a particular class has (the number of classes which extend it), this metric very clearly maps to the number of projects that utilize the one we are examining.
It turns that in a closed system, like an application which you calculate CK metrics for, this is easy to calculate.
For an open system, like a project hosted by Apache, this is much harder.

We toyed with a number of methods for arriving at this metric, including:
\begin{enumerate}
  \item The number of downloads
  \item The number of references
\end{enumerate}

Our original idea of using the number of downloads as an indication of popularity was reasonable, but the data was hard to come by.
In addition, it didn't accurately reflect this metric as projects which didn't depend on a particular project might still download that project if another included it.
This would result in certain projects getting hugely bloated number; for example, HTTPClient, which is a component of the HTTPComponents library, is only included by a few projects.
If we look at the number of downloads of this artifact, however, we see a massive increase in the number of downloads as HTTPComponents gets downloaded quite frequently as dependency of the Play framework.

So that leads us to trying to count the number of references.
One could imagine doing this by looking at the meta-data available through the Maven repository.
One option would be to scan all projects in Maven for projects that list this as a dependency; we can not reasonable do this with the resources available to us.
It would require mirroring many Maven repositories, trying to scan through them, and aggregating that data.

Thankfully, someone had already done this.
They published their data to the web, so we were able to download it and use it for our experiment.
We would like to extend immense gratitude to Dr. Fernando Rodriguez Olivera, who created and mainted the web resource https://mvnrepository.com/ which allowed us to examine this metric with very solid data \cite{olivera2017mvnrepository}.

\subsection{Coupling between Object Classes}

The coupling between object classes makes a lot of sense when you're talking about classes; they should be aware of each other, and know how to call methods on each others objects.
When we're talking about libraries, this is less desirable.
A well written library should not depend on the implementation of it's children to dictate its functionality; therefore, we want this value to be as close to 0 as possible.
There are a few cases (paired libraries) where we might be able to envision some cyclic dependencies, but that will most likely be an exception rather than a rule.

During our initial investigation into the DIT metric, we came across a number of cyclic dependencies.
After switching to a more simple approach, described above, the number of cycles greatly reduced; this suggests that there may be an interesting pattern to how projects add and remove dependencies.
This investigation was not done while writing this paper, as it has a series of challenging and interesting problems which would need to be addressed, including:
\begin{enumerate}
	\item Scaling; during some of our runs, we ended up with 30000 projects to explore. Creating a graph of this size would require some work
	\item Caching; to do this in a reasonable timeframe, we would benefit from a local mirror of the repositories. Without this, it was taking days to calculate the metric from a single project
\end{enumerate}

\subsection{Response for a Class}

This value was meant to be a count of the possible function calls that could result from a function on the class being called; that is, explore all code paths in each public function and count the number of function calls.
We can easily translate this to talk about systems if we accept just a first-step measure of function calls.
This value then becomes the number of unique methods calls exposed to applications that depend on this one.

We ended up using the Java disassembler to get the list of public functions for a class, and aggregate a list of the number of functions that would be called for those public functions.

\subsection{Lack of Cohesion of Methods}

Chidambar and Kemerer original meant this to be the count of the set of methods not used by a pair of classesl.
The exact meaning has come under fire in the past, and several replacements have been proposed.
For the Interactive Complexity (IC)-metric suite, we propose that this value represent the number of unused imports within a project.
The logic for this is simple; if I include numpy just to multiply two arrays, I'm not making use of all the features it provides, and therefore it should not be as heavily weighted as my including of another library which my entire application is based upon.

Initially, we had calculated this metric using the Soot framework; however, this was found to be rather slow.
Soot was intended for a different type of project, and was overkill for the analysis we were doing here; instead, we opted to use the Maven dependency plugin again, which had a feature to find the unused dependencies within a project.

\section{Case Study}

Our case study utilized the Software Metrics Framework \cite{software_metrics_framework} to analyze a large number of projects.
There are two possible approaches: in the synchronic approach, one compares projects of similar size, or uses some normalization factor to make them comparable.
This normalization is required to account for the fact that very large projects, regardless of interactive complexity, will have larger number of bugs than small projects.
In the diachronic approach, we compare the metric in one project as it changes over time, to the number of bugs as they change over time.
Essentially that is normalizing the project to itself.
We decided to use this diachronic approach for several reasons.
First, it solves the problem of finding the normalization factor (lines of code? Some other complexity metric?).
Second, the synchronic approach requires that one make a decision about which version of the software is used: the first version? The latest version? Third, by sampling the code at every marked release, we have many data points, so that even if this is far from the final version, we can determine a general trend.
Each release is a version of the code that was deemed of sufficient quality to be given to the public, and will have a list of faults there fixed in that version.

We analyzed 80 projects, which had a cumulative 1600 versions, 1000 of which were tagged in the version control system and would compile.
These projects were randomly selected from the Apache foundation's selection, so long as they matched the following criteria:

\begin{itemize}
  \item They must have a Git mirror
  \item They must use JIRA for issue tracking
  \item They must be Java projects that use Maven
\end{itemize}

SMF determines these attributes for us by checking the Apache JIRA installation for the projects presence, and the source repository for a pom.xml file.

As a test of our assumptions above, we examine the number of lines of code (LOC) as a metric.
There was indeed a very strong correlation to fault rate, suggesting that lack of normalization would lose the interactive complexity “signal” in the noise of project size effects.
LOC has been previously dismissed a poor measure of software complexity \cite{rosenberg_misconceptions_1997}.
It has been speculated the LOC is a good measure of the scale of a project; and for that reason, the larger the project, the more bugs it will have \cite{rosenberg_misconceptions_1997}.
Using LOC instead of our time interval as a synchronic normalizer would be an interesting experiment, and a possible avenue of further research.

Figure \ref{project_list} shows a complete list of selected projects,along with the median values for their metrics across all versions.
We believe this to be a reasonable sample size for statistical analysis, but believe that it is worth selecting a few representatives projects to provide concrete examples will add value to the discussion.
The criteria for these projects are:
\begin{itemize}
	\item They must have at least 10 versions which we could identify in the codebase
    \item 80\% of the versions analyzed must have successfully compiled; without this, only a subset of our metrics could be applied
    \item They must have non-zero counts of reported and fixed bugs in each version; without this, we would have nothing meaningful to analyze.
\end{itemize}

\begin{figure}[!htb]
	\label{accumulo}
	\centering
	\includegraphics[scale=0.75]{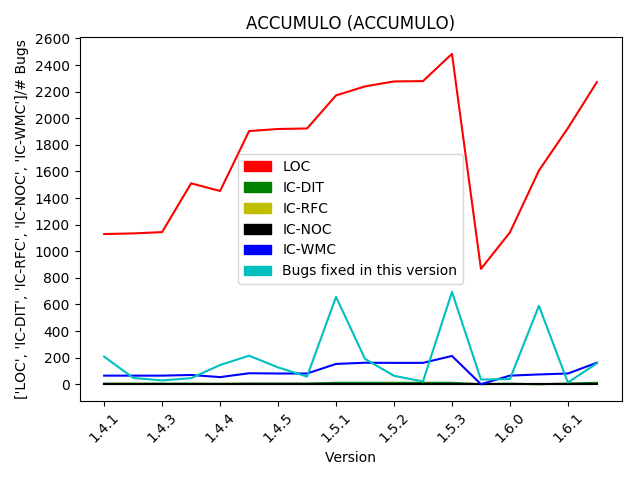}
	\caption{Apache Accumulo}
\end{figure}

\begin{figure}[!htb]
	\label{httpclient}
	\centering
	\includegraphics[scale=0.75]{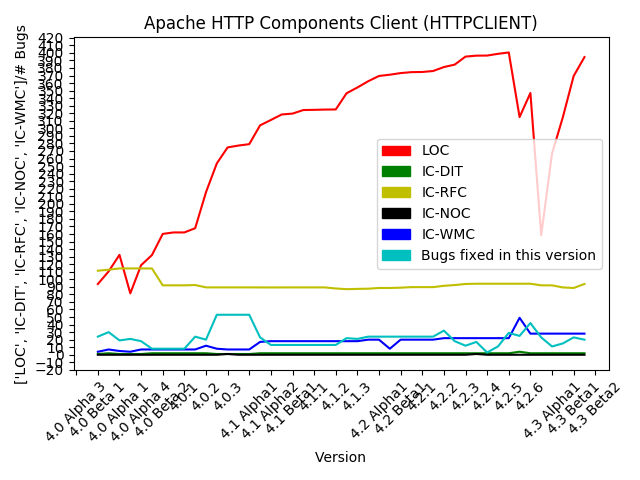}
	\caption{Apache HTTPComponents Client}
\end{figure}

\begin{figure}[!htb]
	\label{rave}
	\centering
	\includegraphics[scale=0.75]{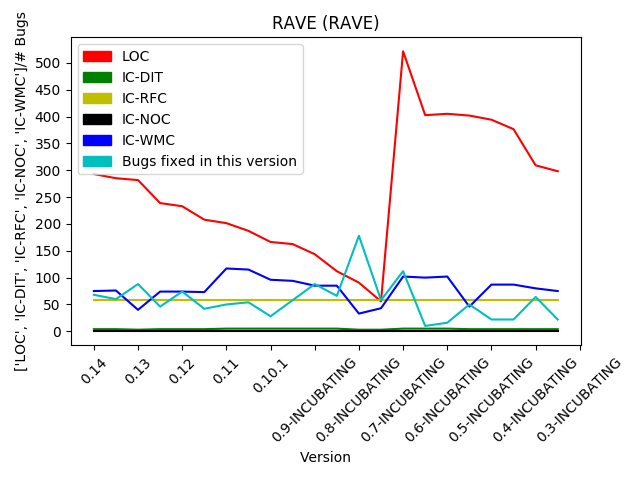}
	\caption{Apache Rave}
\end{figure}

\begin{figure}[!htb]
	\label{cxf}
	\centering
	\includegraphics[scale=0.75]{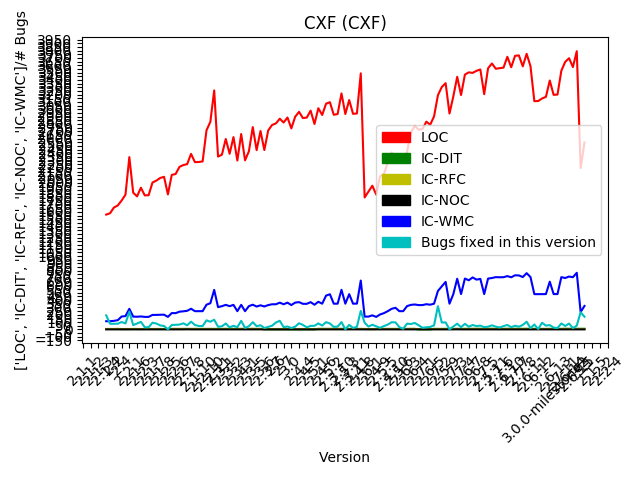}
	\caption{Apache CXF}
\end{figure}

\begin{figure}[!htb]
	\label{log4j2}
	\centering
	\includegraphics[scale=0.75]{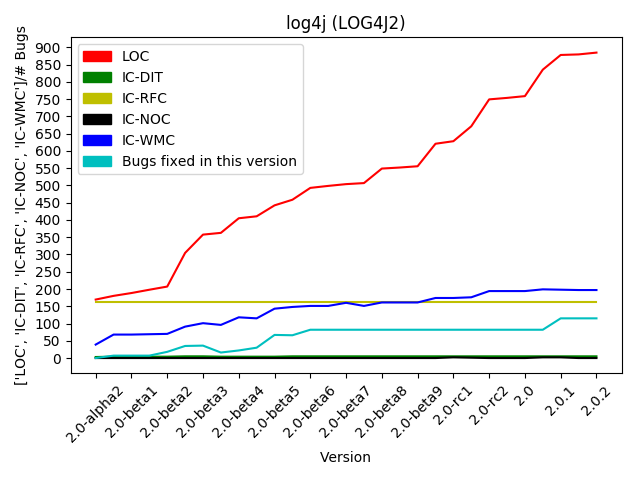}
	\caption{LOG4J2}
\end{figure}

The projects that were selected as representative are the Apache Accumulo, Apache HTTPComponents Client library, Apache Rave, Apache CXF, and log4j2.
Figures \ref{accumulo}, \ref{httpclient}, \ref{rave}, \ref{cxf}, and \ref{log4j2} show the calculated metrics for these projects, alongside the bug fix count, for each versions that we were able to match to a commit in their VCS.

\subsection{Problems Encountered}

During this case study, a number of issues were encountered.
Some of them required changes to the initial planned scope of the case study.

\begin{enumerate}
  \item Docker VM was slow; initial hope was to ease the use of SMF by using Docker
    \begin{itemize}
      \item Moved out of Docker, and run on bare metal
    \end{itemize}
  \item Exponentional growth of parent tree when exploring DIT
    \begin{itemize}
      \item Use techniques from web graphs; limit number of explored parents
      \item Limiting number of explored parents doesn't lower quality due to the "6 degrees of seperation" theory
    \end{itemize}
  \item Slow pom.xml scanning
    \begin{itemize}
      \item Wrote custom Maven plugin to perform multiple scans after one pass of pom.xml
    \end{itemize}
  \item Rate limiting of Maven repositories
    \begin{itemize}
      \item Add handling of SIGINT to delay project execution
    \end{itemize}
\end{enumerate}

\section{Results and Interpretation}

For reference, the metrics we are examining are:
\begin{enumerate}
  \item IC-NOC: Number of Children; number of projects that depend on this one
  \item IC-DIT: Depth of the Inheritance Tree; longest chain of parent dependencies
  \item IC-LCOM1: Lack of Cohesion Methods; imports unused within the project, but included anyway
  \item IC-WMC: Weighted methods per class; number of libraries that this one depends on
  \item IC-RFC: Response For Class; number of publically exposed methods
\end{enumerate}

Figure \ref{combined_pearsons} shows the correlation and p-value of each metric across all projects; of great note is the low p-values of IC-NOC, IC-RFC and IC-LCOM1.

Figure \ref{p_values} shows the correlation and p-value of each metric across each project.

\begin{center}
\captionof{figure}{List of p-values}%
\begin{longtable}{| p{.40\textwidth} | p{.40\textwidth} | p{.20\textwidth} |}
	\hline
	\textit{Metric} & \textit{Correlation} & \textit{two-tailed p-value} \\
    \hline
	IC-NOC & -0.0897 & 0.060 \\
	IC-DIT & 0.0585 & 0.155 \\ 
	IC-LCOM1 & 0.195 & 0.014 \\
	IC-WMC & 0.0189 & 0.643 \\ 
	IC-RFC & 0.598 & 2.15e-17 \\
  LOC & 0.439 & 2.027e-31 \\
	\hline
\end{longtable}
\addtocounter{table}{-1}%
\label{combined_pearsons}
\end{center}

Overall, we see very good p-values for IC-NOC, IC-RFC and IC-LCOM1.
A positive correlation in IC-RFC and IC-LCOM1 suggests that as these values increase, the number of bugs fixed also increase.
Likewise, the negative correlation for IC-NOC suggests that as more projects begin to rely on this one, the number of bugs fixed in a release goes down.

These results are the opposite of what we expected.
We had thought that as the number of children went up, it would be the result of a good project (hence, fewer bugs).
Additionally, as the number of publicly exposed functions go up, we would expect the project quality to go down (more bugs) due to the nature of encapsulation.

The results of IC-LCOM1 could have gone either way; including libraries to utilize one feature, if that feature is well designed, may reduce the complexity of the project we are examining.
These results suggest that doing this (a high LCOM means there are many unused dependencies) does increase the number of bugs fixed in a release.

\begin{figure}[!htb]
        \label{dit_bugs_vs_pval}
        \centering     
        \includegraphics[scale=0.75]{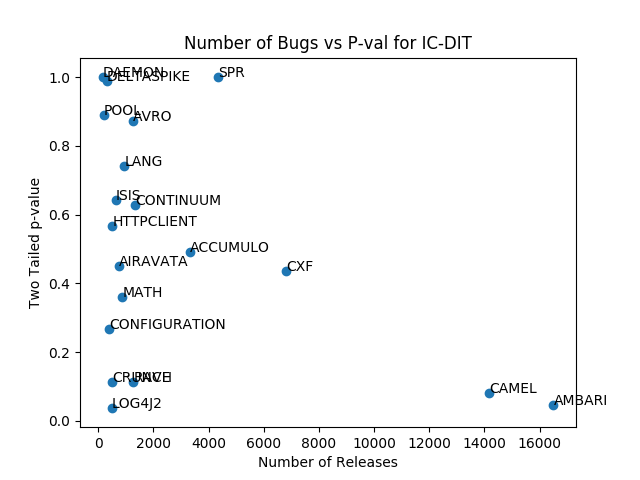}      
        \caption{IC-DIT Number of Bugs vs pval}                      
\end{figure}

\begin{figure}[!htb]
        \label{dit_releases_vs_pval}
        \centering     
        \includegraphics[scale=0.75]{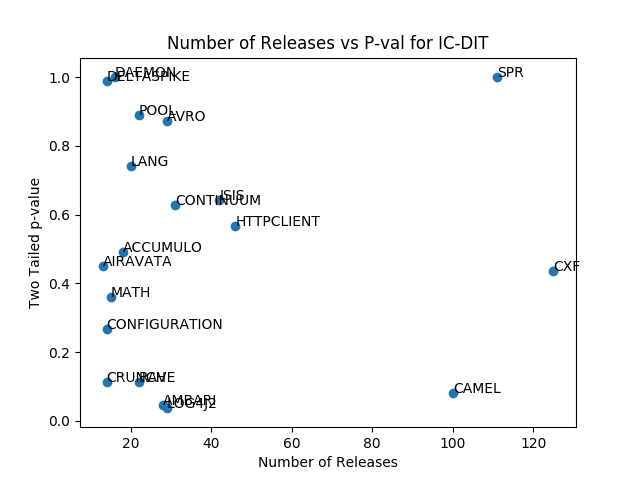}  
        \caption{IC-DIT Number of Releases vs pval}                          
\end{figure}

\begin{figure}[!htb]
        \label{dit_X_vs_pval}
        \centering     
        \includegraphics[scale=0.75]{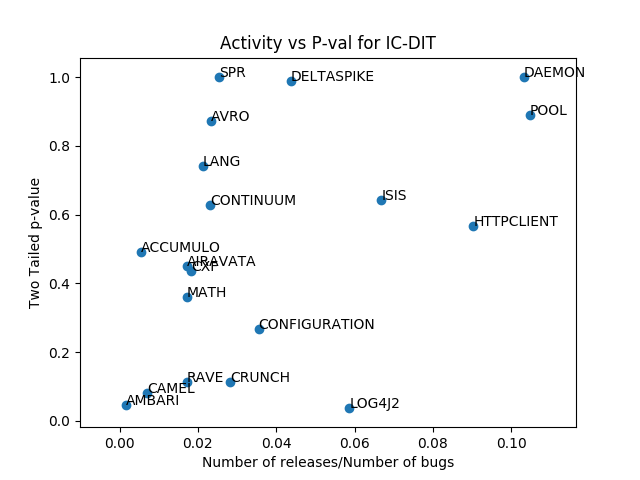}
        \caption{IC-DIT Activity vs pval}
\end{figure}

\begin{figure}[!htb]
        \label{lcom1_bugs_vs_pval}
        \centering     
        \includegraphics[scale=0.75]{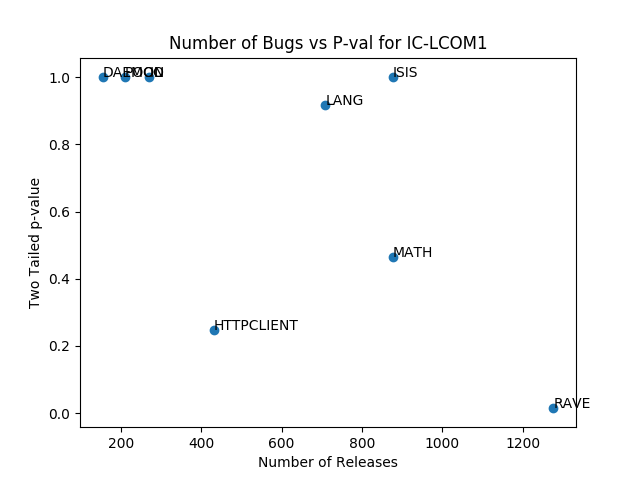}
        \caption{IC-LCOM1 Number of Bugs vs pval}
\end{figure}

\begin{figure}[!htb]
        \label{lcom1_releases_vs_pval}
        \centering     
        \includegraphics[scale=0.75]{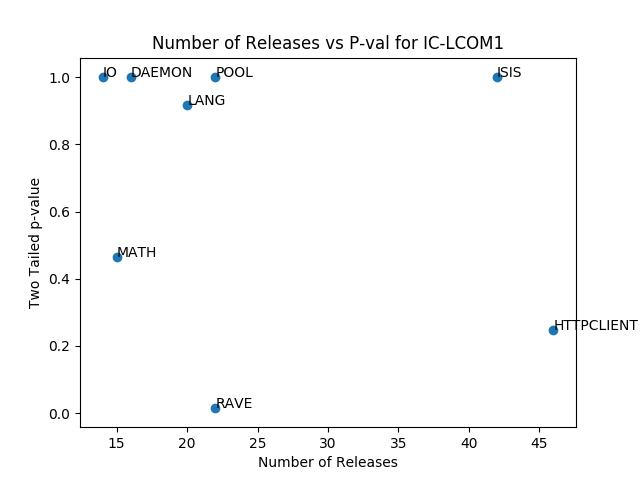}
        \caption{IC-LCOM1 Number of Releases vs pval}
\end{figure}

\begin{figure}[!htb]
        \label{lcom1_X_vs_pval}
        \centering     
        \includegraphics[scale=0.75]{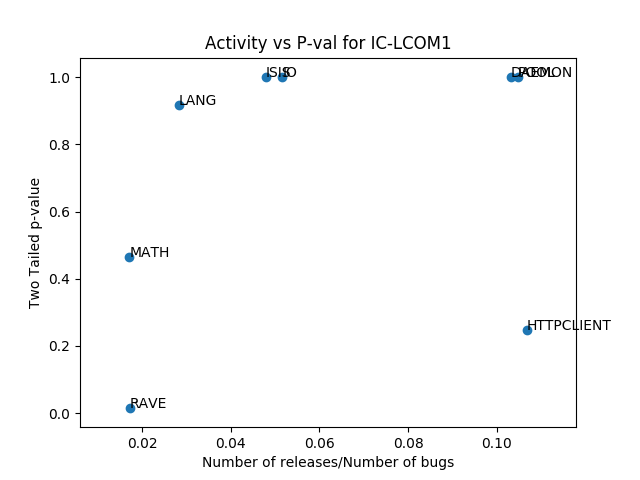}
        \caption{IC-LCOM1 Activity vs pval}
\end{figure}

\begin{figure}[!htb]
        \label{rfc_bugs_vs_pval}
        \centering     
        \includegraphics[scale=0.75]{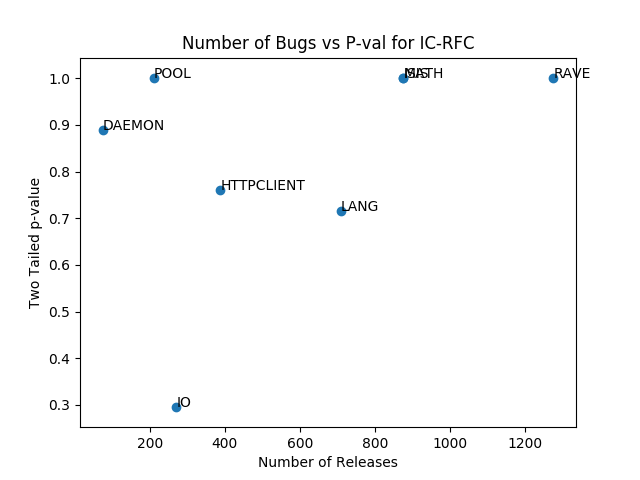}
        \caption{IC-RFC Number of Bugs vs pval}
\end{figure}

\begin{figure}[!htb]
        \label{rfc_releases_vs_pval}
        \centering     
        \includegraphics[scale=0.75]{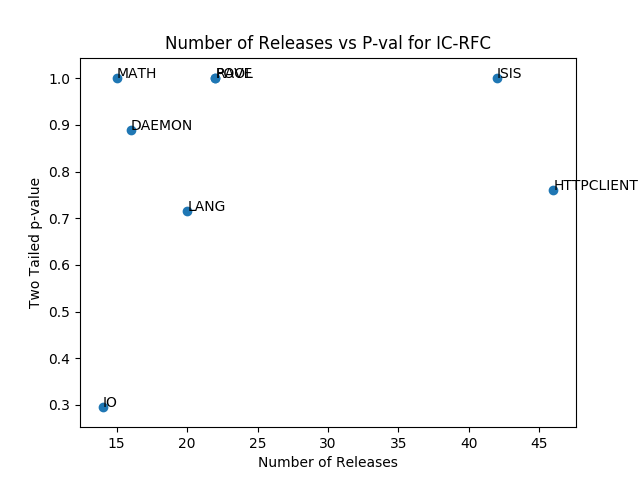}    
        \caption{IC-RFC Number of Releases vs pval}                        
\end{figure}

\begin{figure}[!htb]
        \label{rfc_X_vs_pval}
        \centering     
        \includegraphics[scale=0.75]{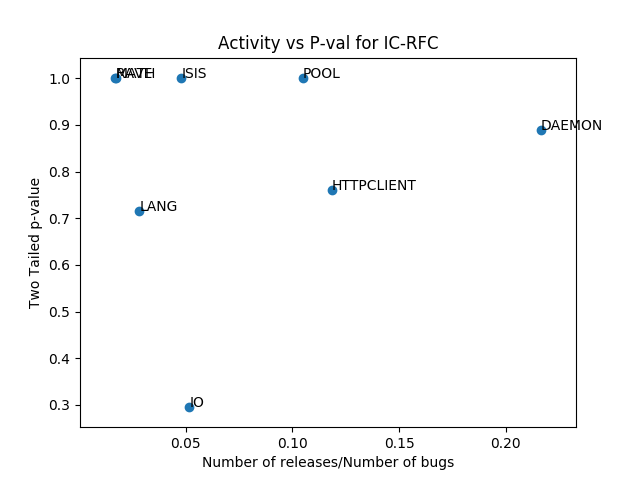}
        \caption{IC-RFC Activity vs pval}
\end{figure}

\begin{figure}[!htb]
        \label{wmc_bugs_vs_pval}
        \centering     
        \includegraphics[scale=0.75]{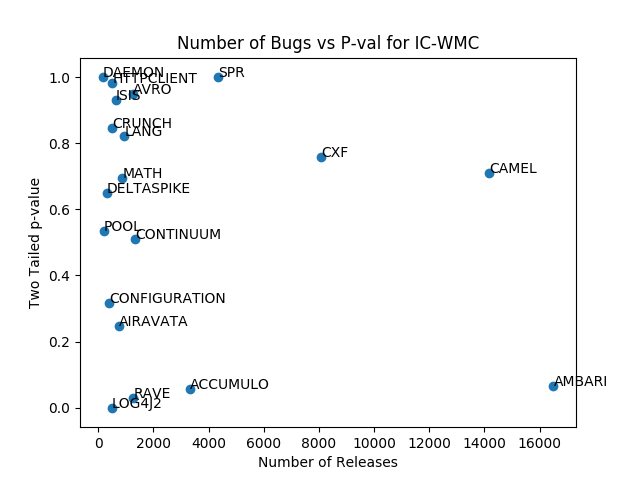}
        \caption{IC-WMC Number of Bugs vs pval}
\end{figure}

\begin{figure}[!htb]
        \label{wmc_releases_vs_pval}
        \centering     
        \includegraphics[scale=0.75]{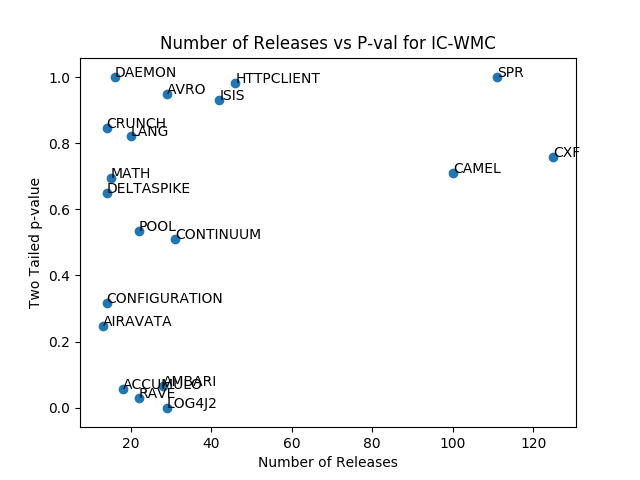}
        \caption{IC-WMC Releases of Bugs vs pval}
\end{figure}

\begin{figure}[!htb]
        \label{wmc_X_vs_pval}
        \centering     
        \includegraphics[scale=0.75]{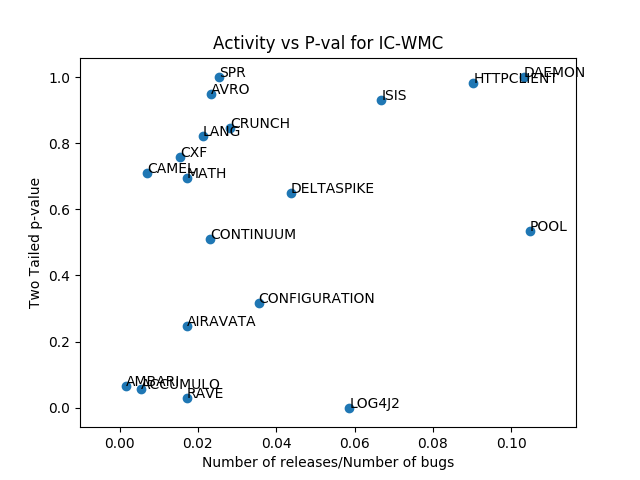}
        \caption{IC-WMC Activity vs pval}
\end{figure}

Despite some strong correlations, we see a huge variation of correlation between a metric and the bug count between projects.
In figures \ref{dit_bugs_vs_pval}-\ref{wmc_X_vs_pval}, which show the p-value mapped against either the bug count or the number of release, we do not see anything that looks like a trend.
However, some really large values have a really strong correlation, and some really small values have a really weak correlation.
based on this evidence, we started hypothesizing what it might be which would demarcate the projects that had a strong correlation, and the projects which had a weak correlation.

The projects with strong correlation of IC-DIT:
\begin{itemize}
\item LOG4J2
\item CAMEL
\item AMBARI
\end{itemize}

We found that if one took the number of releases divided by the number of bugs, we ended up with a graph where all the projects that had a strong correlation were grouped into the bottom left corner; this means that we can identify those projects for which these metrics have meaning, and possibly ignore the others.

As to why this metric is valuable for some projects, but not others, can be explained by looking at the operation we did to single out these projects.
Number of releases divided by number of bugs, notable the inverse of the number of bugs per release, gives us a good indication of how long the project has been around, and how active the development has been.
We believe that only projects of sufficient scale will exhibit behavior consistent with our analyses; therefore, only those projects will have this correlation.
In the graphs from figures \ref{dit_bugs_vs_pval}-\ref{wmc_X_vs_pval} this value is what "Activity" is referring too.
To verify this, we will need to run more analysis on larger projects.

Below, we present a number of graphs.
Figures \ref{accumulo} trough \ref{log4j2} show the complexity of projects over a period of time.
The blue line represents the number of bugs fixed in each version.
Our goal is to find correlations between this blue line and all the other lines; the good news is that we see a correlation in many of these projects.
The bad news is that some projects don't have this correlation, which leads us to the remainder of the graphs.

Figures \ref{dit_bugs_vs_pval}, \ref{lcom1_bugs_vs_pval}, \ref{rfc_bugs_vs_pval}, \ref{wmc_bugs_vs_pval} show the number of totals bugs in a project versus the confidence level of our metrics being correlated to the number of bugs.
Each graph shows the confidence of a single metric, as described in their titles.
There does not seem to be a correlation, beyond a few projects that are where we expect (Camel, Ambari).
Note the great vertical spread.

Figures \ref{dit_releases_vs_pval}, \ref{lcom1_releases_vs_pval}, \ref{rfc_releases_vs_pval}, \ref{wmc_releases_vs_pval} show the number of releases a project has versus their confidence level for a given metric.
We seem a much greater spread head, not quite as horizontal.

For the remainder of the figures, \ref{dit_X_vs_pval}, \ref{lcom1_X_vs_pval}, \ref{rfc_X_vs_pval}, \ref{wmc_X_vs_pval}, we tooks the number of releases and divided it by the number of bugs.
The previous two types of graphs is what led us to do this; neither of them was perfect, but each had some value.
Using this method, we can see a much stronger correlation between the confidence level and the value.
In some of these graphs, LOG4J2 is mis-classified due to the way bugs are counted by their project managers and by the way it was put into SMF.
They would put all bugs towards a release identifier, but not tag that release in the source code; this meant that none of those releases with many bugs had metric calculations.

\section{Conclusion}

Our findings indicate a strong connection between interactive complexity and fault rate.
This is consistent with previous literature (cite things here) which demonstrate a strong connection between the CK metric suite and fault rate.

Future works should explore whether or not this connection can be narrowed to speciific parts of the application.
This could help developers narrow down bugs, and improve the state of the art for many software projects.

\printbibliography

\pagebreak

\begin{center}
	\centering
	\captionof{table}{List of Projects and Values}%
  \begin{longtable}{|p{.20\textwidth} | p{.10\textwidth} | p{.10\textwidth} | p{.10\textwidth} | p{.10\textwidth} | p{.12\textwidth} | p{.08\textwidth} |}
	\hline
    \textit{Project} & \textit{IC-DIT} & \textit{IC-WMC} & \textit{IC-RFC} & \textit{IC-LCOM1} & \textit{LOC} & \textit{Bugs} \\
	\hline
    \endhead
	NUMBERS &  &  &  &  &  & 6 \\
BAHIR &  &  &  &  &  & 36 \\
CODEC & 1.00 & 1.50 &  & 8.00 & 13547.00 & 90 \\
COMPRESS & 1.50 & 2.50 &  & 8.00 & 18300.50 & 182 \\
IMAGING &  &  &  &  &  & 90 \\
DBUTILS & 1.50 & 3.00 & 195.00 & 10.00 & 4774.50 & 26 \\
FILEUPLOAD & 2.00 & 5.00 & 233.00 & 8.00 & 6068.50 & 59 \\
CONFIGURATION & 3.00 & 30.00 &  & 20.00 & 49677.50 & 273 \\
NET & 1.00 & 1.00 & 0.00 & 0.00 & 25016.00 & 225 \\
OGNL &  &  &  &  &  & 134 \\
POOL & 1.00 & 1.00 & 0.00 & 0.00 & 12874.50 & 152 \\
CRUNCH & 6.00 & 147.50 & 8896.50 & 41.50 & 44190.00 & 352 \\
LAUNCHER & 0.00 & 0.00 &  &  & 2575.00 & 3 \\
VFS & 1.00 & 10.00 & 0.00 & 0.00 & 29694.00 & 167 \\
HCATALOG &  &  & 0.00 & 0.00 & 30105.00 & 291 \\
HTTPCLIENT & 2.00 & 18.00 & 8961.00 & 17.00 & 64382.00 & 502 \\
LOG4J2 & 5.00 & 151.00 & 16177.00 & 63.00 & 100754.00 & 296 \\
LANG & 1.00 & 2.50 & 4597.00 & 10.00 & 56617.50 & 469 \\
PROXY & 2.00 & 13.00 & 0.00 & 0.00 & 4206.00 & 2 \\
BCEL & 0.00 & 0.00 &  & 11.00 & 29818.50 & 146 \\
CXF & 6.00 & 342.00 &  &  & 559973.00 & 7053 \\
OMID &  &  & 0.00 & 0.00 & 20769.00 & 12 \\
JXPATH & 2.00 & 11.00 & 1430.00 & 10.00 & 26342.00 & 56 \\
VALIDATOR & 1.00 & 6.00 & 0.00 & 0.00 & 20376.00 & 85 \\
DIGESTER & 0.00 & 0.00 & 608.50 & 8.00 & 8264.50 & 41 \\
JCI & 2.50 & 14.00 & 330.00 & 19.00 & 5575.50 & 16 \\
JSPWIKI &  &  & 0.00 & 0.00 & 110956.00 & 385 \\
AMBARI & 4.00 & 20.00 &  & 8.00 & 1153514.00 & 14049 \\
BEANUTILS & 1.00 & 4.00 &  & 9.00 & 34836.00 & 150 \\
DBCP & 4.00 & 27.00 & 1264.50 & 11.00 & 17781.00 & 192 \\
ACCUMULO & 6.00 & 81.00 &  & 56.00 & 382276.00 & 2772 \\
BIGTOP & 0.00 & 0.00 &  & 0.00 & 136731.00 & 1373 \\
CURATOR & 4.00 & 48.00 & 1582.00 & 32.00 & 32440.00 & 202 \\
SPR & 0.00 & 0.00 &  &  & 494552.00 & 3115 \\
CHAIN & 0.00 & 0.00 &  & 9.00 & 15681.00 & 31 \\
CONTINUUM & 2.00 & 9.00 &  &  & 80980.00 & 1010 \\
RNG & 3.00 & 9.00 & 0.00 & 0.00 & 9890.00 & 2 \\
EMAIL & 4.00 & 19.00 & 259.00 & 14.00 & 8306.50 & 48 \\
TRANSACTION &  &  &  &  &  & 11 \\
ABDERA & 4.00 & 75.50 &  & 78.50 & 491355.00 & 135 \\
COCOON & 0.00 & 0.00 &  &  & 1294129.00 & 347 \\
TEXT & 2.00 & 4.00 & 0.00 & 0.00 & 13881.00 & 136 \\
DISCOVERY & 0.00 & 0.00 & 263.00 & 8.00 & 4027.00 & 9 \\
RAVE & 4.00 & 82.50 & 5814.00 & 66.00 & 52045.50 & 332 \\
RAT &  &  &  &  &  & 62 \\
DAEMON & 1.00 & 1.00 & 51.00 & 8.00 & 21827.00 & 115 \\
ISIS & 0.00 & 0.00 & 0.00 & 0.00 & 366280.50 & 433 \\
GERONIMO &  &  & 0.00 & 0.00 & 10035.00 & 270 \\
SCXML & 2.00 & 18.00 & 0.00 & 0.00 & 17367.00 & 91 \\
JEXL & 1.00 & 3.00 &  & 8.00 & 13859.00 & 107 \\
CSV & 2.00 & 4.00 &  & 10.00 & 5915.00 & 60 \\
FUNCTOR &  &  &  &  &  & 1 \\
DELTASPIKE & 8.00 & 222.00 & 3497.00 & 346.00 & 82394.50 & 308 \\
DIRECTMEMORY & 2.50 & 94.50 &  &  & 33673.00 & 31 \\
COLLECTIONS & 0.00 & 0.00 &  & 9.00 & 53878.50 & 225 \\
CLI & 1.00 & 1.50 &  & 8.00 & 7060.00 & 126 \\
CAMEL & 6.00 & 331.50 &  &  & 642452.50 & 6226 \\
WHISKER &  &  &  &  &  & 1 \\
WINK &  &  &  &  & 128976.00 &  \\
AIRAVATA & 7.00 & 164.00 &  &  & 150093.00 & 825 \\
SLIDER & 3.00 & 59.00 & 0.00 & 0.00 & 116581.00 & 442 \\
MATH & 1.00 & 2.00 & 9806.00 & 8.00 & 162806.00 & 510 \\
MRQL &  &  &  &  &  & 8 \\
LOGGING &  &  & 0.00 & 0.00 & 6632.00 & 77 \\
BEAM & 11.00 & 318.00 &  & 53.00 & 175955.00 & 889 \\
JCS &  &  &  &  &  & 97 \\
language\_tool &  &  &  &  &  &  \\
CRYPTO & 2.00 & 5.00 &  &  & 9451.00 & 32 \\
BSF & 1.50 & 7.50 &  &  & 3571.50 & 16 \\
DDLUTILS & 2.00 & 33.00 & 1257.00 & 16.00 & 40198.00 & 105 \\
BATIK & 4.00 & 41.00 &  & 13.00 & 225067.00 & 25 \\
BVAL & 3.00 & 21.00 &  & 37.00 & 19239.00 & 91 \\
ANY23 & 8.00 & 134.50 &  & 60.00 & 60641.00 & 103 \\
TENTACLES &  &  &  &  &  &  \\
AVRO & 4.00 & 67.00 &  & 0.00 & 132250.00 & 653 \\
CHUKWA & 4.00 & 87.00 &  &  & 78760.00 & 204 \\
WEAVER & 2.00 & 22.50 & 0.00 & 0.00 & 9021.50 & 3 \\
EXEC & 1.00 & 1.00 & 180.00 & 8.00 & 4545.00 & 21 \\
OOZIE & 7.00 & 188.00 &  &  & 139008.00 & 944 \\
IO & 1.00 & 1.00 & 964.00 & 8.00 & 20548.50 & 158 \\

	\hline
	\end{longtable}
\addtocounter{table}{-1}%
\label{project_list}

\noindent (80 rows) \\
\end{center}

\pagebreak

\begin{center}
  \captionof{table}{List of p-values}%
	\begin{longtable}{| p{.30\textwidth} | p{.30\textwidth} | p{.20\textwidth} | p{.20\textwidth} |}
	\hline
		\textit{Project Name} & \textit{Metric} & \textit{Correlation} & \textit{two-tailed p-value} \\
	\hline
    \endhead
HTTPCLIENT & IC-DIT & -0.13 & 0.57 \\
HTTPCLIENT & IC-WMC & -0.01 & 0.98 \\
HTTPCLIENT & IC-LCOM1 & -0.26 & 0.25 \\
HTTPCLIENT & IC-RFC & -0.07 & 0.76 \\
HTTPCLIENT & IC-NOC & -0.09 & 0.66 \\
HTTPCLIENT & LOC & -0.07 & 0.72 \\
SPR & IC-DIT & nan & 1.0 \\
SPR & IC-WMC & nan & 1.0 \\
SPR & LOC & 0.13 & 0.28 \\
LOG4J2 & IC-DIT & 0.63 & 0.04 \\
LOG4J2 & IC-WMC & 0.94 & 0.0 \\
LOG4J2 & IC-NOC & 0.7 & 0.02 \\
LOG4J2 & LOC & 0.89 & 0.0 \\
ACCUMULO & IC-DIT & 0.17 & 0.49 \\
ACCUMULO & IC-WMC & 0.46 & 0.06 \\
ACCUMULO & IC-NOC & nan & 1.0 \\
ACCUMULO & LOC & 0.38 & 0.12 \\
AIRAVATA & IC-DIT & 0.23 & 0.45 \\
AIRAVATA & IC-WMC & 0.35 & 0.25 \\
AIRAVATA & IC-NOC & nan & 1.0 \\
AIRAVATA & LOC & 0.15 & 0.63 \\
AMBARI & IC-DIT & -0.38 & 0.05 \\
AMBARI & IC-WMC & -0.35 & 0.07 \\
AMBARI & LOC & 0.25 & 0.2 \\
AVRO & IC-DIT & -0.03 & 0.87 \\
AVRO & IC-WMC & -0.01 & 0.95 \\
AVRO & IC-NOC & nan & 1.0 \\
AVRO & LOC & -0.09 & 0.64 \\
CAMEL & IC-DIT & -0.18 & 0.08 \\
CAMEL & IC-WMC & -0.04 & 0.71 \\
CAMEL & IC-NOC & nan & 1.0 \\
CAMEL & LOC & 0.03 & 0.79 \\
CONFIGURATION & IC-DIT & 0.32 & 0.27 \\
CONFIGURATION & IC-WMC & 0.29 & 0.32 \\
CONFIGURATION & LOC & 0.34 & 0.23 \\
DAEMON & IC-DIT & nan & 1.0 \\
DAEMON & IC-WMC & nan & 1.0 \\
DAEMON & IC-LCOM1 & nan & 1.0 \\
DAEMON & IC-RFC & 0.05 & 0.89 \\
DAEMON & IC-NOC & -0.45 & 0.16 \\
DAEMON & LOC & -0.01 & 0.97 \\
IO & IC-LCOM1 & nan & 1.0 \\
IO & IC-RFC & 0.35 & 0.3 \\
IO & IC-NOC & nan & 1.0 \\
IO & LOC & 0.4 & 0.16 \\
LANG & IC-DIT & -0.08 & 0.74 \\
LANG & IC-WMC & -0.06 & 0.82 \\
LANG & IC-LCOM1 & 0.03 & 0.92 \\
LANG & IC-RFC & 0.11 & 0.72 \\
LANG & IC-NOC & -0.16 & 0.57 \\
LANG & LOC & 0.1 & 0.67 \\
CONTINUUM & IC-DIT & -0.1 & 0.63 \\
CONTINUUM & IC-WMC & -0.14 & 0.51 \\
CONTINUUM & IC-NOC & nan & 1.0 \\
CONTINUUM & LOC & -0.18 & 0.33 \\
CRUNCH & IC-DIT & 0.46 & 0.11 \\
CRUNCH & IC-WMC & -0.06 & 0.85 \\
CRUNCH & IC-NOC & nan & 1.0 \\
CRUNCH & LOC & 0.07 & 0.81 \\
CXF & IC-DIT & 0.07 & 0.44 \\
CXF & IC-WMC & -0.03 & 0.76 \\
CXF & IC-NOC & 0.01 & 0.95 \\
CXF & LOC & -0.07 & 0.44 \\
DELTASPIKE & IC-DIT & -0.0 & 0.99 \\
DELTASPIKE & IC-WMC & 0.13 & 0.65 \\
DELTASPIKE & IC-NOC & nan & 1.0 \\
DELTASPIKE & LOC & -0.27 & 0.34 \\
MATH & IC-DIT & 0.26 & 0.36 \\
MATH & IC-WMC & 0.12 & 0.69 \\
MATH & IC-LCOM1 & 0.21 & 0.46 \\
MATH & IC-RFC & nan & 1.0 \\
MATH & IC-NOC & -0.29 & 0.34 \\
MATH & LOC & 0.02 & 0.95 \\
POOL & IC-DIT & 0.03 & 0.89 \\
POOL & IC-WMC & 0.14 & 0.53 \\
POOL & IC-LCOM1 & nan & 1.0 \\
POOL & IC-RFC & nan & 1.0 \\
POOL & IC-NOC & -0.21 & 0.42 \\
POOL & LOC & -0.11 & 0.64 \\
ISIS & IC-DIT & 0.09 & 0.64 \\
ISIS & IC-WMC & -0.02 & 0.93 \\
ISIS & IC-LCOM1 & nan & 1.0 \\
ISIS & IC-RFC & nan & 1.0 \\
ISIS & LOC & -0.2 & 0.24 \\
RAVE & IC-DIT & -0.35 & 0.11 \\
RAVE & IC-WMC & -0.46 & 0.03 \\
RAVE & IC-LCOM1 & -0.51 & 0.02 \\
RAVE & IC-RFC & nan & 1.0 \\
RAVE & IC-NOC & nan & 1.0 \\
RAVE & LOC & -0.32 & 0.15 \\

	\hline
\end{longtable}

  \noindent (80 rows) \\
  \addtocounter{table}{-1}%
  \label{p_values}
\end{center}


\end{document}